\documentclass[pdflatex,sn-mathphys-ay]{sn-jnl}

\usepackage{graphicx}%
\usepackage{multirow}%
\usepackage{amsmath,amssymb,amsfonts}%
\usepackage{amsthm}%
\usepackage{mathrsfs}%
\usepackage[title]{appendix}%
\usepackage{xcolor}%
\usepackage{textcomp}%
\usepackage{manyfoot}%
\usepackage{booktabs}%
\usepackage{algorithm}%
\usepackage{algorithmicx}%
\usepackage{algpseudocode}%
\usepackage{listings}%
\usepackage{tabularray}
\usepackage{graphicx}  
\usepackage{tabularx}
\usepackage{adjustbox} 
\usepackage{makecell}
\usepackage{cleveref}

\usepackage{ragged2e}     
\usepackage{booktabs}     

\theoremstyle{thmstyleone}%
\theoremstyle{thmstyletwo}%

\theoremstyle{thmstylethree}%

\begin{document}

\title[Light-weighted foundation model for seismic data processing based on representative and non-redundant pre-training dataset]{Light-weighted foundation model for seismic data processing based on representative and non-redundant pre-training dataset}


\author*[1]{\fnm{Xintong} \sur{Dong}}\email{18186829038@163.com}

\author[1]{\fnm{Wenshuo} \sur{Yu}}\email{yuwenshuo1998@163.com}

\author[1]{\fnm{Jun} \sur{Lin}}\email{lin\_jun@jlu.edu.cn}

\author[2]{\fnm{Zhenbo} \sur{Guo}}\email{guozhenbo01@cnpc.com.cn}

\author[1]{\fnm{Hongzhou} \sur{Wang}}\email{hzwang21@mails.jlu.edu.cn}

\author[1]{\fnm{Jianhao} \sur{Yang}}\email{yangjh1722@mails.jlu.edu.cn}

\affil*[1]{College of Instrumentation and Electrical Engineering, Jilin University, Changchun 130026, Jilin, 
China}

\affil[2]{Bureau of Geophysical Prospecting (BGP), CNPC, Zhuozhou 072751, China}


\abstract{In the fields of computer vision (CV) and remote sensing (RS), foundational models typically follow the "big data + large model parameters" paradigm. However, the application of this strategy in seismic data processing faces several challenges: seismic data is difficult to obtain and the scarcity of publicly available datasets make it difficult to construct large-scale datasets. Additionally, the high computational cost associated with a large number of model parameters restricts widespread research in this domain. Therefore, we propose a lightweight seismic processing foundational model paradigm (SPFM), which aims to overcome the limitations of traditional methods by data engineering and network architecture innovation. Specifically, we propose an innovative dataset construction strategy that generates more seismic data by data augmentation techniques, including collecting publicly available field data and using generative diffusion models (GDM) for data enhancement. Furthermore, we optimize the data distribution by employing dimensionality reduction, cluster analysis, and stratified sampling methods, reducing redundant information while preserving important seismic features, thus constructing a comprehensive dataset. In terms of network architecture design, we introduce the selective structured state-space model (Mamba) structure, which effectively captures global features of seismic data and alleviates the quadratic growth of computational complexity inherent in Transformer-based models, thereby improving computational efficiency. This model, pre-trained with only four A800 GPUs, outperforms traditional methods across multiple tasks, including denoising, interpolation, frequency-band extrapolation, and resolution enhancement. The lightweight paradigm provides an efficient solution for intelligent seismic data processing, significantly advancing the generalization capability and accessibility of seismic data processing.}




\maketitle

\section{Introduction}\label{sec1}

In recent years, foundation models have achieved breakthrough progress in computer vision (CV) and natural language processing (NLP), demonstrating exceptional feature learning and transfer capabilities in multiple tasks. Studies have shown that foundation models can effectively extract general representations by large-scale data pretraining and adapt efficiently to downstream tasks, significantly enhancing data processing and analysis performance as well as generalization ability [\cite{lu2024visual}, \cite{bai2023benchmarking}, \cite{hong2024spectralgpt}]. However, in the field of geophysics, research on foundation models remains in the exploratory stage, and the relevant technological framework has yet to be established.

Although deep learning techniques have seen some application in geophysics in recent years [\cite{liu2022seismic}, \cite{dong2020denoising}, \cite{kaur2021seismic}, \cite{zhao2022unsupervised}, \cite{wang2024efgw}], research on foundation models is still in its early stages. Currently, only a limited number of studies have attempted to introduce foundation models into geophysics. For instance, \cite{harsuko2022storseismic} proposed the StorSeismic framework, which adopts a data-driven strategy, integrating neural network pretraining and fine-tuning while incorporating the bidirectional encoder representations from transformers (BERT) attention mechanism to enhance the representation of key geometric features in seismic data and achieve multi-task processing capabilities. \cite{sheng2024seismic} further proposed a development workflow for geophysical foundation models and built the seismic foundation model (SFM) based on a self-supervised learning pretrained transformer architecture, achieving significant results in multiple seismic processing and interpretation tasks. \cite{si2024seisclip} developed the SeisCLIP model, which utilizes a contrastive learning strategy to pretrain on multimodal seismic waveform spectra and their corresponding local and global event information, generating transferable seismic data representations. This model has demonstrated superior performance in seismic event classification, source localization, and source mechanism analysis. However, most of these studies directly transfer model architectures from other fields without fully considering the unique characteristics of seismic data. For example, seismic data are often not publicly available, making it challenging to construct large-scale training sets [\cite{dong2022seismic}, \cite{wang2021seismogen}]. Moreover, compared to the field of computer vision, computational resources in geophysics are relatively limited, constraining the deployment of high-computation-demand foundation models in this field [\cite{liu2024foundation}].

\begin{figure*}
\centering
\includegraphics[width=5in, keepaspectratio]{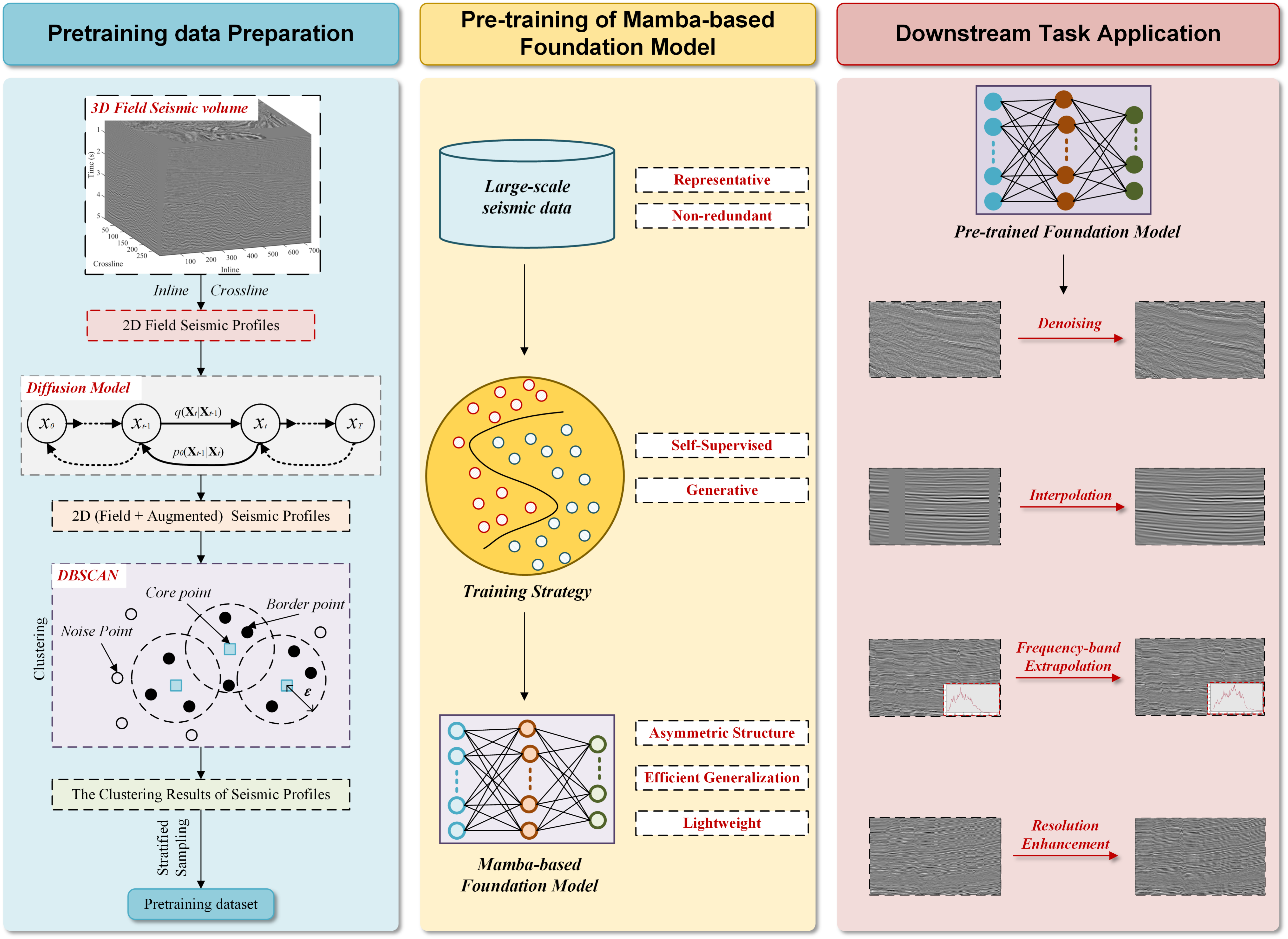}\\
\caption{The proposed SPFM development workflow: construction of a comprehensive dataset, design of a lightweight foundation model, and downstream task applications. First, by acquiring a large number of 2D seismic profile and using diffusion models for data augmentation, combined with dimensionality reduction, clustering, and stratified sampling strategies, a comprehensive pre-training dataset was constructed. Next, a self-supervised learning strategy was applied to pre-train the dataset, and the proposed lightweight foundation model was built and trained. Finally, by downstream tasks such as denoising, interpolation, frequency-band extrapolation (low-frequency extrapolation), and resolution enhancement, the multifunctionality of SPFM in seismic data processing is fully demonstrated.}
\label{fig:fig1}
\end{figure*}

To address these challenges, we propose a comprehensive seismic data processing workflow by developing a novel lightweight foundation model (SPFM) to mitigate the constraints of seismic data collection and computational resources, as illustrated in Fig. \ref{fig:fig1}. First, we collect publicly available three-dimensional (3D) seismic data and employ a sliding window approach to extract two-dimensional (2D) seismic profiles in both inline and crossline directions. Based on these 2D seismic profiles, we proposed a data augmentation method using a generative diffusion model (GDM) to expand the dataset. Next, by integrating clustering and stratified sampling strategies, we select representative and non-redundant data as the training set for the pretraining stage, effectively mitigating data redundancy and imbalanced geological feature distribution issues. Additionally, we also proposed a pre-trained model based on the selective structured state-space model (Mamba) architecture and introduce a self-supervised pretraining strategy, as proposed by \cite{he2022masked}, to enhance network training efficiency. This strategy captures both global and local features of seismic data while avoiding the quadratic computational complexity issues associated with transformer architectures.

To validate the effectiveness of the proposed approach, we apply this model to post-stack seismic profile processing, achieving high signal-to-noise ratio, high fidelity, and high-resolution seismic data processing results. The experiments were conducted using four A800 GPUs for pretraining and fine-tuning, with pretraining taking only 19.50 hours. Experimental results indicate that, compared to existing popular methods, the proposed foundation model exhibits significant advantages in seismic data processing, better adapting to the complexity and diversity of seismic data, thereby advancing research and applications of geophysical foundation models.

\section{Seismic processing foundation model}\label{sec2}

In this section, we will first introduce the pre-training data used and its processing methods. Subsequently, we will elaborate on the framework of SPFM, along with its pre-training and fine-tuning strategies.

\subsection{Training data preparation}\label{subsec1}

\subsubsection{Data collection}\label{subsubsec1}

We first collected a large amount of available 3D seismic volumes. Then, a 128×128×128 3D sliding window was used to perform sliding sampling along the inline, crossline, and depth directions, thereby extracting 3D seismic patches. Subsequently, 2D seismic profiles along the Inline and Crossline directions, each with a size of 128×128, were extracted from these 3D seismic patches. Based on this, seismic patches containing large areas of blank space were removed. After data cleaning, we obtained a total of 100,000 2D seismic profiles, encompassing various geological features such as normal faults, reverse faults, folds, and so on.

\subsubsection{Data augmentation}\label{subsubsec2}

In the pre-training process of seismic processing foundation models, the quantity of training data is crucial for the extraction of geological features of seismic data. Insufficient training data may prevent the model from effectively capturing the complex geological features of the seismic data, thereby affecting its generalization capability. Therefore, ensuring an adequate amount of training data is key to enhancing the model's robustness and adaptability. However, the available seismic data is often limited, making it difficult to meet the large-scale data requirements of foundation models. Diffusion models, as a latent variable structure, learn the underlying structure of a dataset by modeling the diffusion process of data points in latent space. They have been widely applied in tasks such as image generation [\cite{zhu2023conditional}, \cite{ho2022cascaded}], image denoising \cite{zhu2023denoising}, and image inpainting \cite{zhang2024mmginpainting}. Recently, researchers have introduced diffusion models into the field of seismic exploration, including applications like seismic denoising [\cite{sun2025unsupervised}, \cite{peng2024seismic}], seismic interpolation [\cite{wang2024seisfusion}], and seismic velocity synthesis [\cite{wang2024controllable}], achieving remarkable results in these areas. Based on this, we proposed using the denoising diffusion probabilistic model (DDPM) [\cite{ho2020denoising}], as a data augmentation strategy to augment 2D seismic profiles, thereby expanding the dataset required for the pre-training phase and further enhancing the geological representation capability of SPFM.

\begin{figure*}
\centering
\includegraphics[width=5in, keepaspectratio]{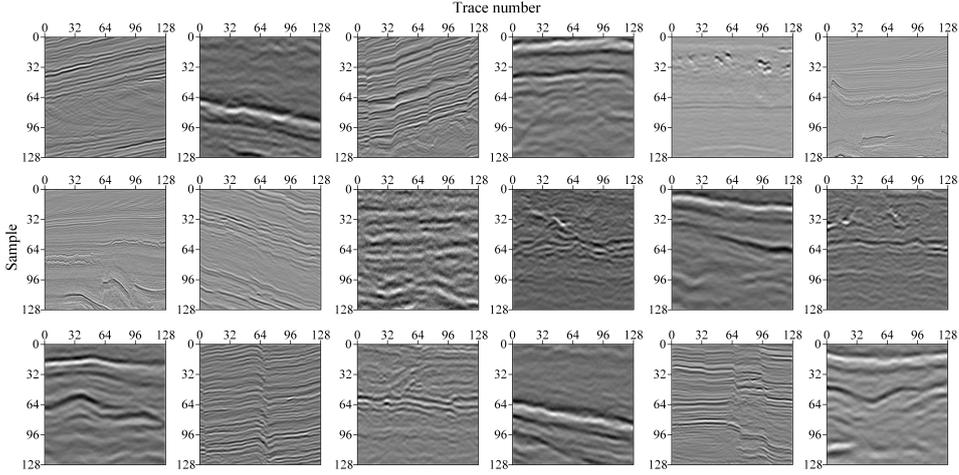}\\
\caption{The augmented data generated by GDM, illustrating its diverse characteristics.}
\label{fig:fig2}
\end{figure*}

\begin{figure*}
\centering
\includegraphics[width=5in, keepaspectratio]{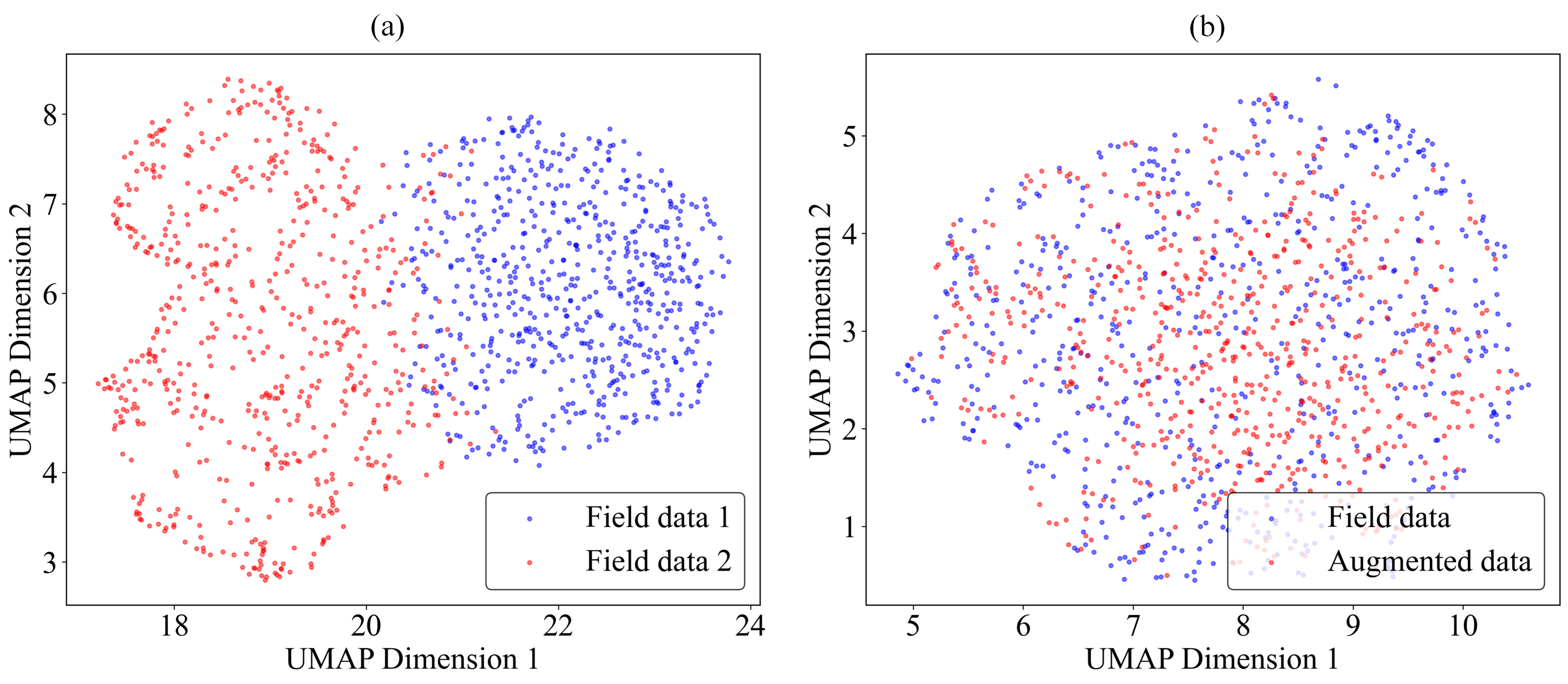}\\
\caption{Dimensionality reduction results of augmented and field data: (a) Dimensionality reduction visualization results of two different survey areas; (b) Dimensionality reduction visualization results of field data and its augmented data.}
\label{fig:fig3}
\end{figure*}

We used the DDPM algorithm to augment the 100,000 collected 2D seismic profiles to 1,000,000. It is important to note that we first divided the 100,000 2D seismic profiles according to the survey area and then performed augmentation separately for each subset. The augmentation results are partially shown in Fig. \ref{fig:fig2}. To validate the consistency between the augmented data and the collected 2D seismic profiles, we applied the uniform manifold approximation and projection (UMAP) algorithm [\cite{mcinnes2018umap}] to reduce the dimensionality of both the augmented and real data. Compared to the t-SNE algorithm [\cite{van2008visualizing}], UMAP demonstrates superior computational efficiency in large-scale seismic data dimensionality reduction due to its lower computational complexity and efficient nearest-neighbor graph construction. The dimensionality reduction results are shown in Fig. \ref{fig:fig3}. It can be observed that the sampled points of seismic data from different survey areas do not overlap in the dimensionality-reduced space, indicating a significant difference in their probability distributions. The sampled points of the seismic data augmented using the DDPM method closely overlap with those of the original field seismic data in the dimensionality-reduced space, further validating the consistency in the probability distributions between the augmented data and the field seismic data.

\subsubsection{Data filtering}\label{subsubsec3}

By the data augmentation strategy based on DDPM, we have obtained a large number of 2D seismic profiles. However, these data contain numerous similar and redundant geological features, and their spatial distribution is uneven, resulting in a lack of representativeness. Directly using these data for pre-training would not only waste computational resources but also significantly increase training time, thereby affecting training efficiency. To effectively address this problem, we propose a data filtering method that combines clustering analysis and sampling strategies. This method aims to select seismic profiles with high representativeness and low redundancy from the augmented dataset, serving as the training set for the pre-training phase, thus improving the model's training efficiency and generalization ability.

\begin{figure*}
\centering
\includegraphics[width=4in, keepaspectratio]{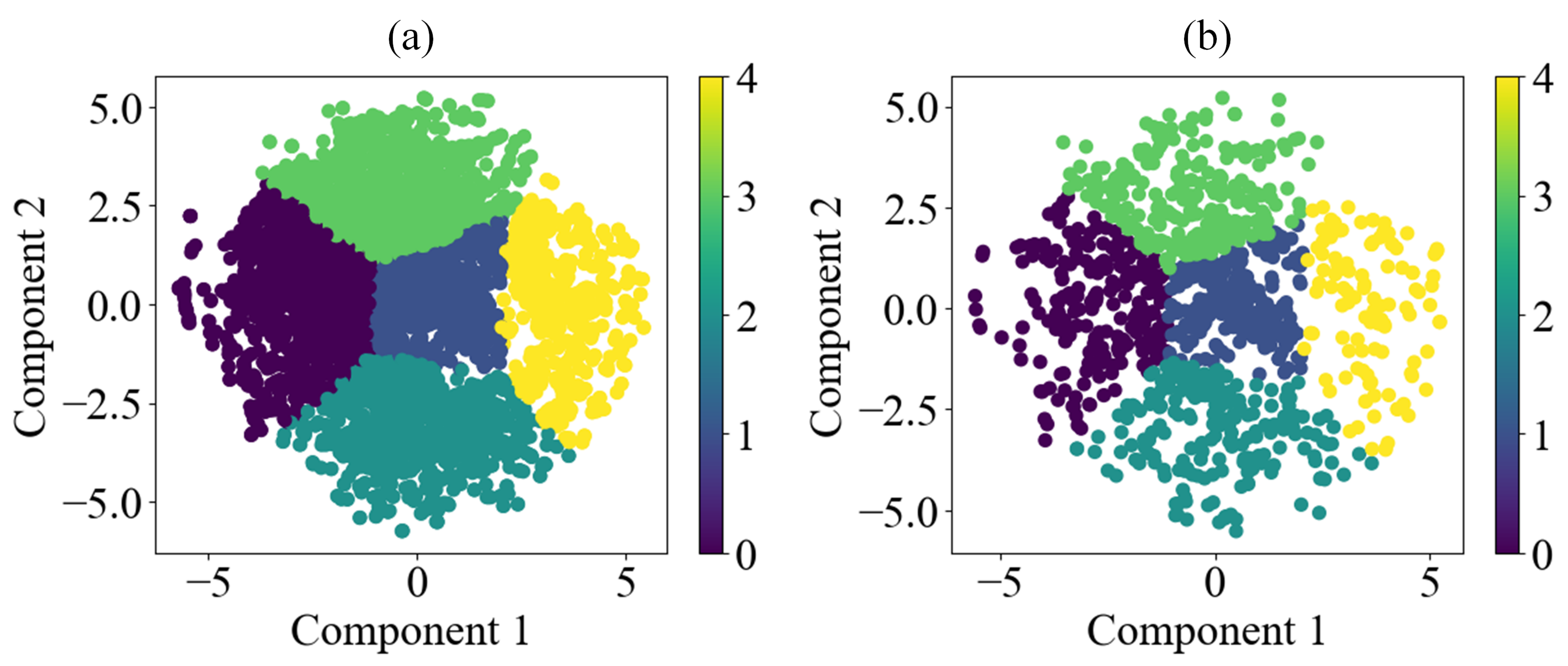}\\
\caption{Dimensionality reduction visualization of data after DBSCAN clustering and stratified sampling. (a) Dimensionality reduction projection of data based on DBSCAN clustering. (b) Dimensionality reduction projection of data after stratified sampling.}
\label{fig:fig4}
\end{figure*}

We adopt the density-based spatial clustering of applications with noise (DBSCAN) [\cite{ester1996density}] method as the clustering strategy for this paper, applied to effectively cluster the augmented seismic profiles. This method is a density-based clustering mechanism that can automatically identify seismic profiles with similar geological features, while effectively removing noise and redundant data.  By this approach, we can perform classification of the seismic profiles, thus providing a more efficient and representative training set for subsequent sampling and model training. After performing data clustering, we used a stratified sampling strategy [\cite{tipton2013stratified}] to select representative data from each cluster. The process is as follows: First, the data is divided into multiple subsets based on the clustering results, with each subset representing a different category. Then, based on the data distribution characteristics and importance of each category, data samples are randomly selected from each subset in proportion to ensure that the selected samples adequately reflect the geological features of each category while reducing redundancy and repetition. The dimensionality reduction visualization results of DBSCAN clustering and stratified sampling are shown in Fig. \ref{fig:fig4}.

\subsection{The pre-training strategy of SPFM}\label{subsec2}

Seismic data exhibits multi-scale characteristics, physical field coupling, and high-dimensional sparsity, which pose special challenges for training foundation models. Unlike natural images, which are represented by discrete pixel intensities, seismic data contains the dynamic features of the elastic wave field, precisely describing the variation of the wave field in space and time, revealing the physical properties of the underground medium and the details of wave propagation. Seismic data not only contains complex reflection structures but also involves dynamic information closely related to the physical properties of stratigraphy (such as impedance, anisotropy, etc.). These features must be recorded and processed with high precision to more accurately describe the propagation laws of the waves. Additionally, seismic profiles demonstrate the lateral continuity and vertical impedance variations of the stratigraphy, which differ from the clear discrete semantic boundaries in natural images. The physical properties in seismic data often exhibit gradual or abrupt patterns, whereas natural images consist of distinct boundaries and discrete semantic units. More importantly, seismic data is constrained by wave equations, with wave propagation dependent on geophysical parameters like anisotropy and attenuation mechanisms in the underground medium. This results in significant differences between the statistical distribution of seismic data and natural images. Therefore, directly transferring the visual foundation model designed for natural images to seismic data processing often struggles to effectively capture the connection between the wave propagation operator and the physical properties of the underground medium. To address these domain differences, we adopted a seismic processing foundation model based on a self-supervised pretraining paradigm [\cite{he2022masked}]. This approach was proposed primarily due to the scarcity of labeled data in practical exploration. In the self-supervised learning framework, contrastive learning typically distinguishes different features using "negative samples." However, in seismic data, some survey areas may have very similar reflection patterns, as they may belong to the same geological feature. This similarity could lead the model to mistakenly treat them as "negative samples," introducing errors and affecting the accuracy of stratigraphic comparison. In contrast, generative methods based on mask reconstruction can automatically learn the wave propagation laws and stratigraphic continuity by restoring missing wavefield data, thereby better processing seismic data.

\subsection{The architecture of SPFM}\label{subsec3}

Most existing foundational models primarily utilize diffusion models or transformer architectures, as both demonstrate significant advantages in handling complex data patterns and capturing long-range dependencies [\cite{kirillov2023segment}, \cite{ko2023meltr}, \cite{he2024lotus}]. However, diffusion models, through their stepwise denoising generation process, can generate high-quality data but face limitations when modeling global information over long time spans. This is particularly problematic for seismic data, where they may fail to fully capture complex spatiotemporal dependencies. On the other hand, transformer effectively extract global features via self-attention mechanisms, making them suitable for seismic data processing. However, their computational complexity grows quadratically with the increase in input size, leading to high computational and memory overheads. Therefore, selecting a lightweight foundational model capable of global feature extraction is especially critical for seismic data processing, to balance computational efficiency while fully leveraging the underlying structure and physical information of the seismic data.

\begin{figure*}
\centering
\includegraphics[width=5in, keepaspectratio]{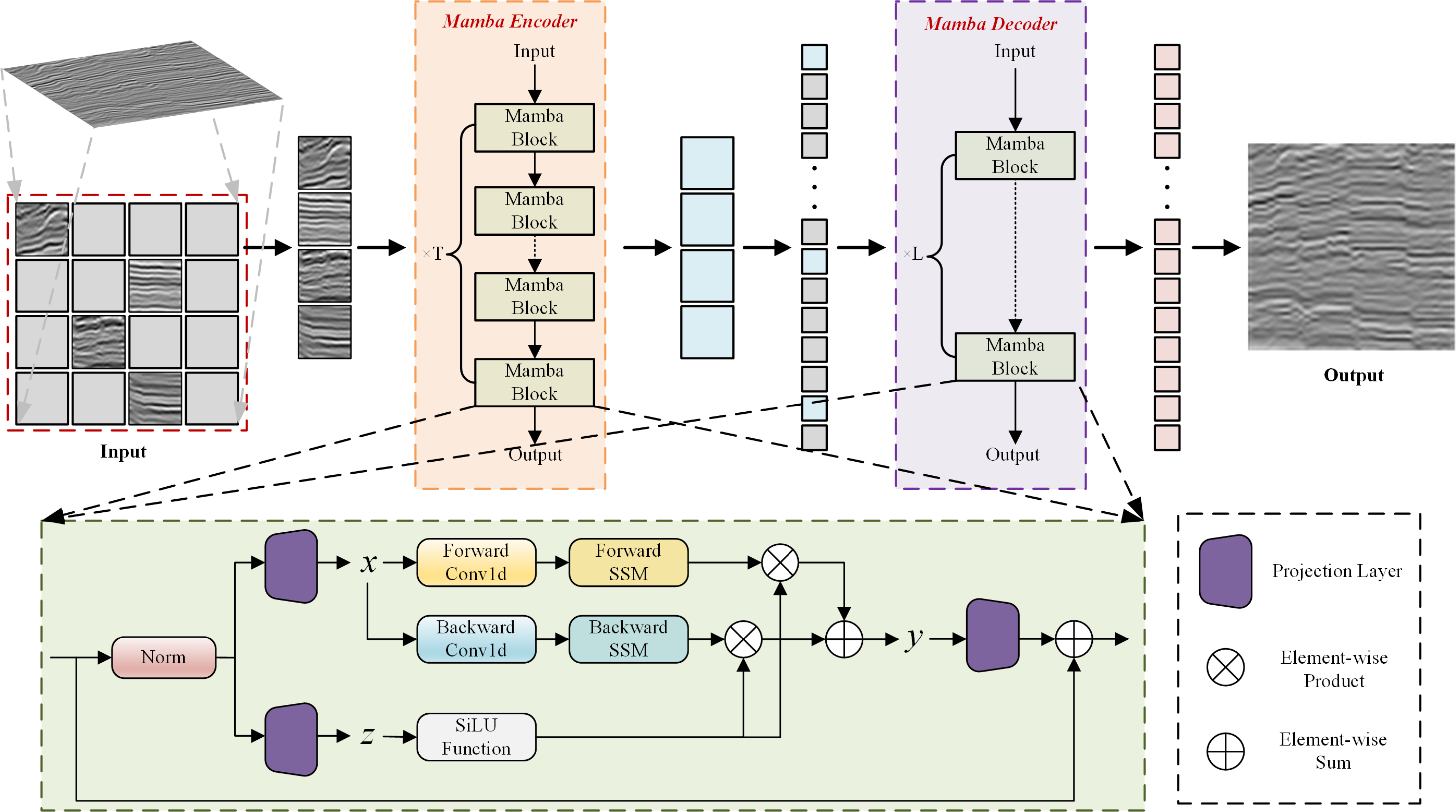}\\
\caption{Pre-training structure of the proposed SPFM. This structure is based on the Mamba framework and utilizes a lightweight network design, allowing for effective extraction of general features of seismic data while significantly reducing computational resource consumption.}
\label{fig:fig5}
\end{figure*}

Mamba is a deep learning network based on state space models [\cite{gu2023mamba}], which extracts global features by the selective scan mechanism while effectively avoiding the quadratic computational complexity associated with the self-attention mechanism in Transformers. This makes it more suitable for seismic data processing. Leveraging this advantage, we adopted Mamba as the network architecture for the proposed SPFM, as shown in Fig. \ref{fig:fig5}. First, the input seismic data is divided into non-overlapping patches of size 16×16. Inspired by [\cite{geng2022multimodal}] and [\cite{bai2023masked}], 75\% of the patches are randomly masked. Then, all the patches are mapped to a higher-dimensional space by linear projection, and positional embeddings are added to the projected vectors. Inspired by the design principles of vision transformer (ViT) and bidirectional encoder representation from transformers (BERT), we introduce an additional class token to represent the global information of the entire sequence of patches. Finally, the class token, along with the projected image patches, forms a new sequence, which is input into the encoder module for processing.

The encoder module consists of twelve Mamba blocks. In each block, the input sequence is first normalized, and then passed by two projection layers to generate two feature sequences. For one of these sequences, feature extraction is performed in both the forward and backward directions, utilizing the SiLU activation function [\cite{elfwing2018sigmoid}] and one-dimensional (1D) convolution operations to capture important features. The resulting features undergo additional processing by a series of linear transformation layers, which incorporate transformation parameters, with the Softplus activation function [\cite{nair2010rectified}] applied to ensure the parameters remain positive. Following this, the features are further computed using a state space model (SSM) to generate the final output. Finally, the forward and backward outputs are combined by the gating mechanism, and a residual connection with the original input is established, ensuring the seamless integration of the original input and the updated features, ultimately producing the encoder module’s final output.

The output of the encoder is fused with the learnable features of the masked patches to obtain a comprehensive feature sequence, which is then input into the decoder. The decoder consists of four Mamba blocks, each further processing the fused features and extracting more information by forward propagation, gradually recovering the missing data patches. Finally, the output of the decoder undergoes a linear transformation to generate the complete seismic image.

\subsection{The fine-tuning strategy of SPFM}\label{subsec4}

\begin{figure*}
\centering
\includegraphics[width=5in, keepaspectratio]{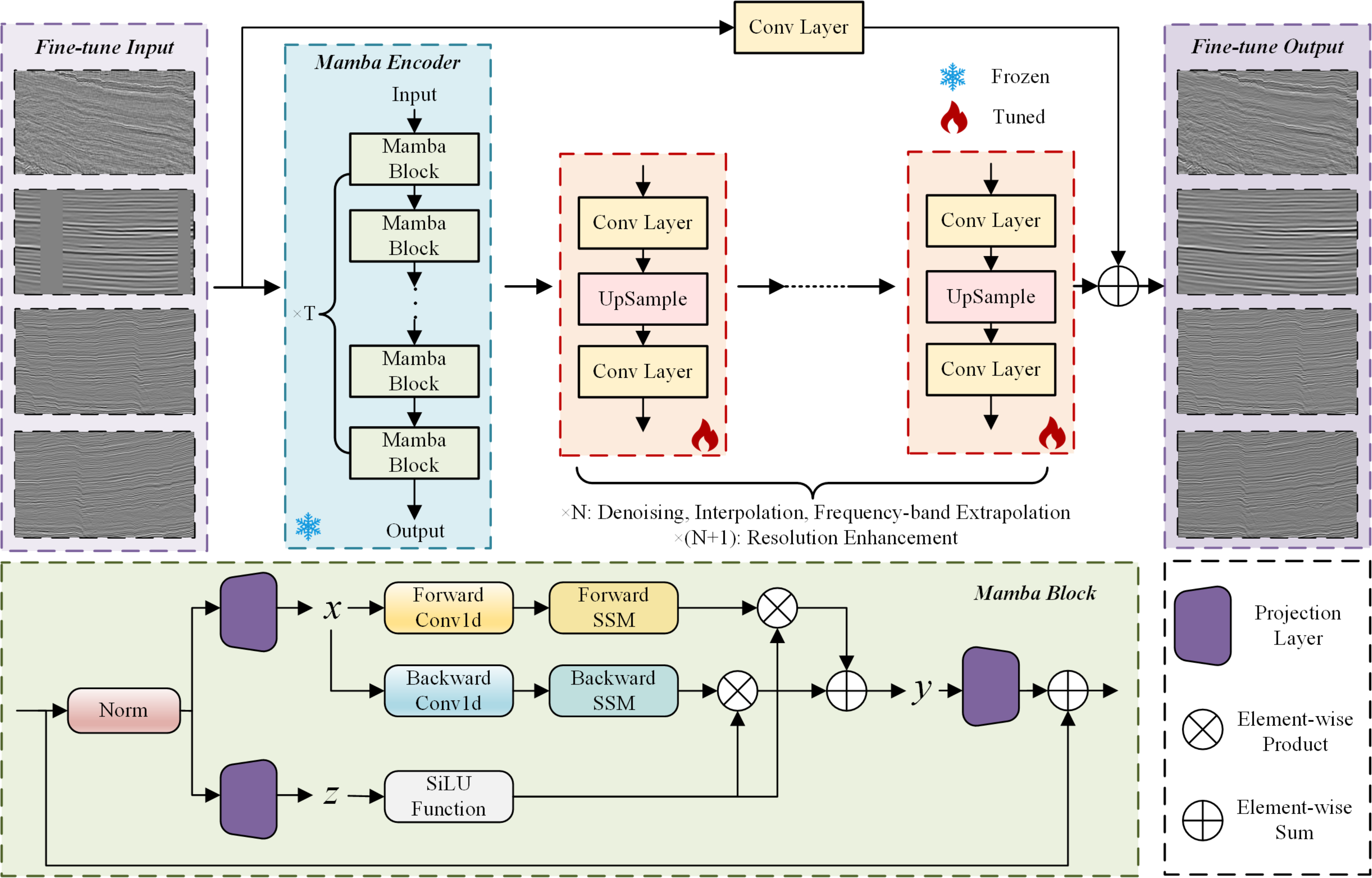}\\
\caption{The fine-tuning structure of the proposed SPFM. The Mamba decoder is replaced with convolution-upsampling-convolution modules, where four-stage convolution-upsampling-convolution modules are used for denoising, interpolation, and frequency-band extrapolation tasks, while five-stage modules are employed for resolution enhancement. Finally, the output of the last convolution layer is element-wise summed with the feature map obtained by convolution operation the input seismic data to generate the final result for downstream tasks.}
\label{fig:fig6}
\end{figure*}

After completing the pretraining of SPFM, we fine-tuned it using field data to enhance the model's adaptability in downstream seismic data processing tasks, such as denoising, interpolation, frequency-band extrapolation (Low-frequency extrapolation), and resolution enhancement. We applied convolution, upsampling, and convolution operations to the output of the last layer of SPFM, repeating this process four times (five times for the resolution enhancement task) to ensure that the feature dimensions remained consistent with the input data. Additionally, we applied convolution operations to the input data and concatenated the processed data with the output from the upsampling module. Finally, a convolutional layer was used to generate the fine-tuned results. The fine-tuning structure of SPFM is shown in Fig. \ref{fig:fig6}.

The pretraining and fine-tuning of the SPFM model are conducted in a Python environment, using a server equipped with four A800-80GB GPUs, a 14-core Intel(R) Xeon(R) Gold 6348 processor, 64GB of memory, running Ubuntu 22.04, Python 3.10, and configured with CUDA 12.1. The hyperparameters in the experimental setup are shown in Table \ref{tab:tab1}.

\begin{table}[h]
\caption{Hyperparameters for different methods}\label{tab:tab1}%
\footnotesize
\begin{tabularx}{\textwidth}{@{} *{6}{>{\centering\arraybackslash}X} @{}}
\toprule
\textbf{Method} & \textbf{Pre-train} & \textbf{Denoising} & \textbf{Interpolation} & \textbf{Frequency-band\newline Extrapolation} & \textbf{Resolution\newline Enhancement} \\
\midrule
Epochs & 1600 & 200 & 200 & 200 & 250 \\
Learning rate & $1 \times 10^{-4}$ & $1 \times 10^{-4}$ & $1 \times 10^{-4}$ & $1 \times 10^{-4}$ & $1 \times 10^{-4}$ \\
Optimizer & AdamW & AdamW & AdamW & AdamW & AdamW \\
Batch size & 9280 & 16 & 16 & 16 & 16 \\
\bottomrule
\end{tabularx}
\end{table}

\section{The principle for solving geophysical underdetermined inverse problems based on deep learning}\label{sec3}

In geophysical exploration, seismic data processing (denoising, interpolation, frequency-band extrapolation and resolution enhancement) is a typical underdetermined inverse problem. Its mathematical expression is as follows: given the observation data matrix $\mathbf{Y} \in \mathbb{R}^{M \times N}$, the goal is to reconstruct the ideal data matrix $\mathbf{X} \in {^{P \times Q}}$, where $M < P$ and/or $N < Q$. The observation process can be modeled by the following equation:

\begin{equation}
\label{eq1}
\mathbf{Y} = \Phi(\mathbf{X}) + \mathbf{N}
\end{equation}

\noindent where $\Phi ( \cdot )$ represents an operator that includes degradation effects such as sparse sampling, frequency band limitations, and noise interference. Traditional processing methods typically rely on several prior assumptions, constructing an objective function by introducing explicit constraints like sparsity and low rank:

\begin{equation}
\label{eq2}
\arg \mathop {\min }\limits_{\mathbf{X}} \left\| {\mathbf{Y} - \Phi(\mathbf{X})} \right\|_2^2 + \lambda R(\mathbf{X})
\end{equation}

\noindent where $R( \cdot )$ is a regularization term, and the regularization parameter   needs to be adjusted empirically. These methods have significant limitations: first, explicit prior assumptions are insufficient to describe the non-stationary features of complex geological structures fully; second, the adjustment of regularization parameters relies on experience and lacks an automatic optimization mechanism; and finally, the computational complexity of iterative optimization algorithms typically grows super-linearly with the data dimension, leading to high computational costs when processing large-scale seismic data.

Deep learning, by a data-driven approach, implicitly learns the prior distribution of geological data $p(\mathbf{X})$, transforming the inverse problem into a function learning task:

\begin{equation}
\label{eq3}
{f_\theta }:\mathbf{Y} \to \mathbf{X}
\end{equation}

Specifically, for the sparse characteristics of the seismic data, we adapted a Mamba architecture based on SSM. The architecture adapts a selective scanning mechanism to effectively capture global features of stratigraphic reflection interfaces. Theoretical analysis shows that the differential equation for the hidden states of the SSM can be expressed as:

\begin{equation}
\label{eq4}
h'(t) = \mathbf{A}h(t) + \mathbf{B}u(t)
\end{equation}

\noindent where this equation has a linear complexity $O(I)$ after discretization, significantly reducing computational costs compared to the transformer model, which has a complexity of $O({I^2})$.

The issue of data representativeness is a key limitation that restricts the generalization ability of deep learning models, especially in seismic data processing. To address this issue, this paper proposes a novel training set construction strategy based on low redundancy and high representativeness. Specifically, we collect publicly available seismic field data, combine it with generative diffusion models for data augmentation, and employ clustering and stratified sampling methods to ensure that the training dataset covers all geological features while minimizing redundancy. The data augmentation process can be expressed by the following stochastic differential equation (SDE):

\begin{equation}
\label{eq5}
\frac{{d\mathbf{X}}}{{dt}} = f(\mathbf{X},t)dt + g(t)dw
\end{equation}

\noindent where $f(\mathbf{X},t)$ is a drift term that governs the evolution of the data over time, and $g(t)$ is a diffusion coefficient.   represents the standard wiener process. This process generates data samples that gradually conform to the target distribution, effectively simulating the spatiotemporal characteristics of seismic data.

We also adopt a pre-training fine-tuning paradigm. In the pre-training phase, the model learns general geological feature representations by minimizing the mean squared error (MSE) loss function, which can be expressed as:

\begin{equation}
\label{eq5}
L = \sum\limits_i {\left\| {{\mathbf{Y}_i} - f({\mathbf{X}_i},\theta )} \right\|_2^2} 
\end{equation}

In the fine-tuning phase, the MSE loss function is still used. By optimizing the prediction results for the target region, the model can quickly adapt to the specific stratigraphic structures of the target region.

\section{Downstream task applications}\label{sec4}

After completing the pretraining of the SPFM, we fine-tuned the model by using post-stack seismic field data to adapt it for downstream task applications, thereby further enhancing its generalization capability. We evaluated the performance of the proposed SPFM on tasks such as seismic denoising, interpolation, frequency-band extrapolation, and resolution enhancement. To achieve this, we fixed the parameters of the encoder, discarded the decoder, and applied a convolution layer, a upsampling layer, and an additional convolution on the output of the last layer of the SPFM encoder. These operations were repeated four times to ensure that the feature dimensions remained consistent with the input data (for the resolution enhancement task, this process was repeated five times). Additionally, we applied convolution operations to the input data and concatenated the results with the output of the upsampling module. Finally, the results were generated by the convolution module. In this section, we demonstrate the performance of the proposed SPFM in different seismic processing tasks by comparing it with post-stack seismic data.

\subsection{Seismic denoising}\label{subsec5}

In seismic exploration, the acquired seismic data is often affected by various random noises. These noises not only significantly reduce the signal-to-noise ratio (SNR) but may also lead to inaccurate imaging of subsurface structures, thereby affecting subsequent seismic interpretation [\cite{anvari2017seismic}, \cite{bakulin2022multiplicative}, \cite{dong2022multiscale}]. Therefore, seismic denoising has become a crucial step in seismic data processing, aiming to effectively suppress noise while preserving effective signals, thereby enhancing the SNR of the data. Due to the difficulty in obtaining paired seismic data with and without noise for the foundational model, we selected data with a high SNR from the Opunake survey in New Zealand as the clean seismic data. Random noise was then added to simulate noisy seismic data. Based on this, we constructed a paired dataset of noisy and clean seismic data for model fine-tuning. In total, 2,000 pairs of seismic data were created for the denoising experiment, including 1,400 pairs for training, 200 pairs for validation, and 400 pairs for testing.

\begin{figure*}
\centering
\includegraphics[width=5in, keepaspectratio]{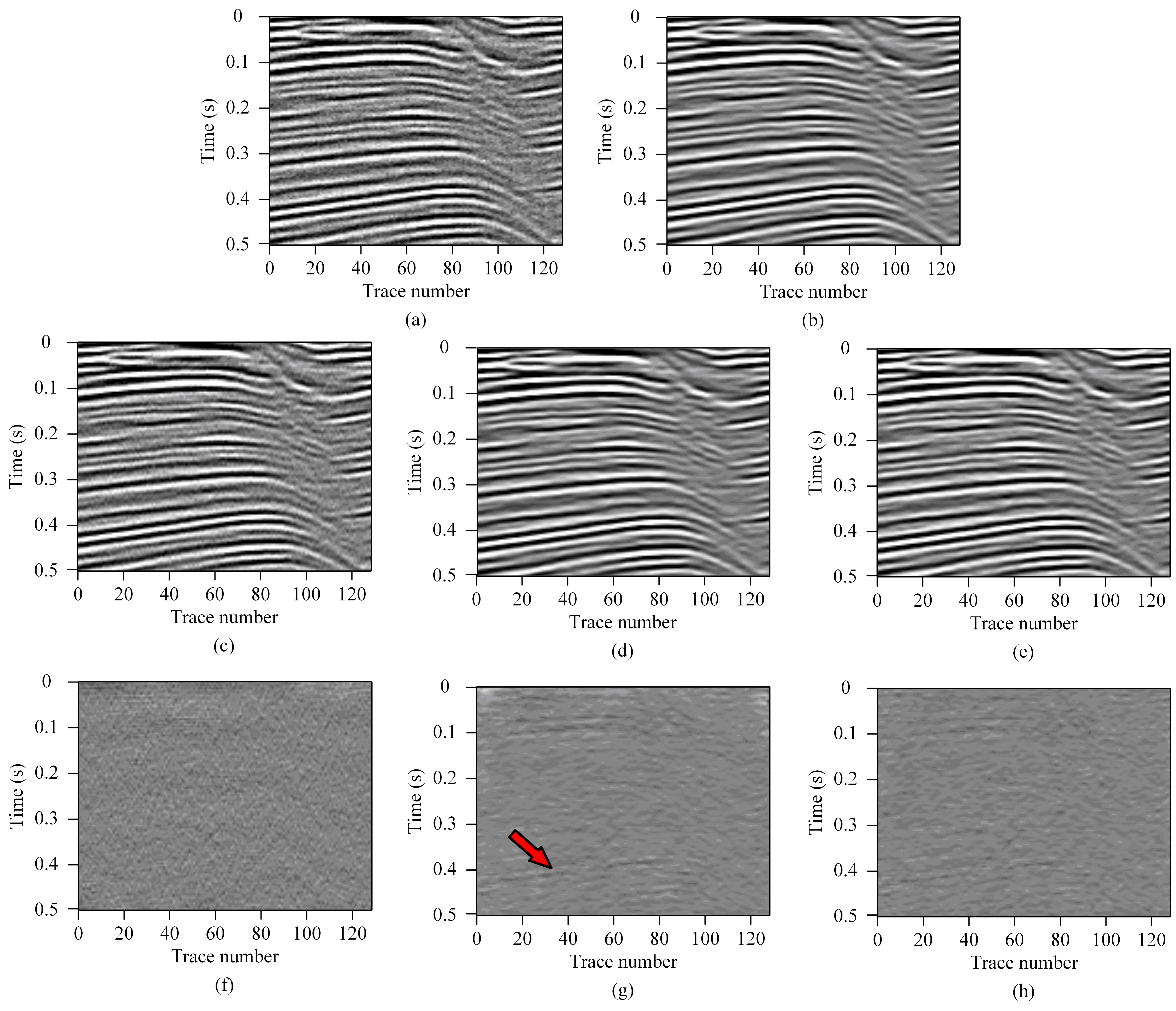}\\
\caption{Comparison of denoising performance of different methods: (a) Noisy seismic data, (b) Clean seismic data, (c)-(d) Denoising results of band-pass filtering, U-Net, and the proposed SFPM method, (f)-(h) Residual results of band-pass filtering, U-Net, and the proposed SFPM method.}
\label{fig:fig7}
\end{figure*}

We conducted a comparative experiment between the proposed SPFM, a traditional method, and a classic deep learning method. The bandpass filter is used as a traditional method for seismic data denoising, which can effectively suppress noise in specific frequency bands while preserving the main features of the seismic signal, thereby significantly improving the SNR. The deep learning method employs the U-Net model for denoising noisy seismic data. The encoder-decoder structure of U-Net is symmetric, allowing it to effectively extract multi-scale features of the seismic data and retain detailed information by skip connections, thus enhancing denoising performance while effectively suppressing noise. The hyperparameters for all comparison methods were optimized, and the experimental results are shown in Fig. \ref{fig:fig7}. After band-pass filtering, a large amount of random noise remains, while U-Net performs better in noise suppression than traditional band-pass filtering. U-Net can effectively remove most of the random noise, but the residual result shows significant signal leakage during the denoising process, indicating certain limitations in global feature extraction, which may lead to the loss of signals (e.g., as indicated by the red arrow). In contrast, the proposed SFPM method not only suppresses random noise more effectively but also better preserves weak signals, achieving superior signal-noise separation.

\subsection{Seismic interpolation}\label{subsec6}

Seismic interpolation is one of the key technologies in seismic data processing. In practical applications, seismic data often suffer from missing traces due to factors such as equipment failure, uneven sampling, or complex topography, which in turn affects the characterization of geological features [\cite{wang2019deep}, \cite{gao2025nersi}]. Missing data can generally be classified into two types: random missing data and continuous missing data. Random missing data refers to traces that are missing in an irregular distribution with uneven intervals, typically caused by equipment failure or noise interference. Continuous missing data, on the other hand, shows a certain continuity in the missing traces, usually due to topographical obstacles or limitations in sampling conditions. Currently, existing seismic interpolation methods can effectively reconstruct missing traces when the proportion of missing data is small. However, when the proportion of missing and continuous traces is large, the complexity of the problem increases significantly, presenting a greater challenge for seismic interpolation. To address this problem, we continue to focus on seismic data from the Opunake survey in New Zealand. By randomly setting some seismic traces and continuous seismic traces to zero, we simulate the missing data scenario, with the missing trace proportion reaching up to 25\%. Based on this, we construct a seismic dataset containing both missing and complete traces for model fine-tuning. In the experiment, a total of 2,000 pairs of seismic data were obtained for denoising experiments, including 1,400 pairs for training, 200 pairs for validation, and 400 pairs for testing.

\begin{figure*}
\centering
\includegraphics[width=5in, keepaspectratio]{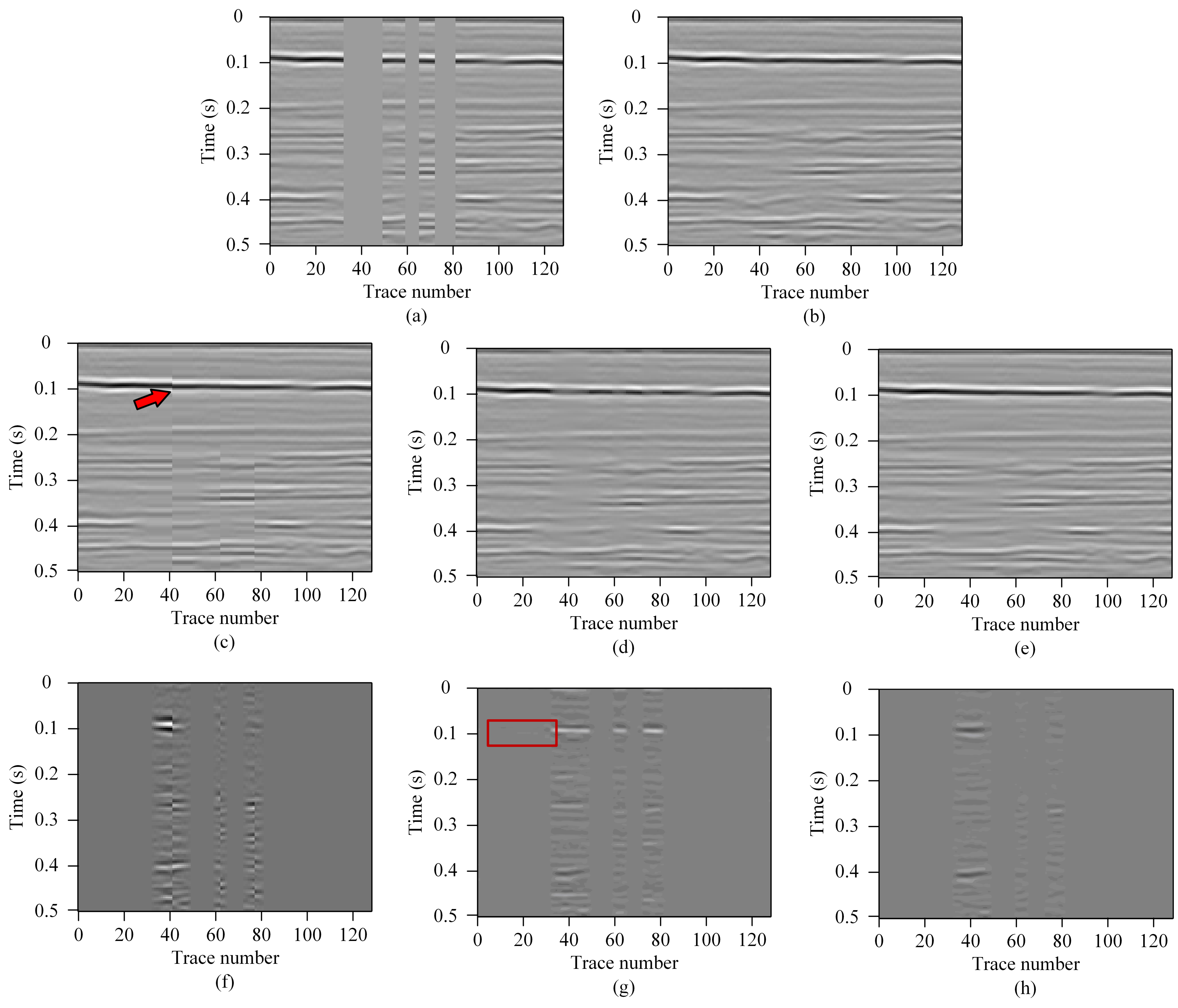}\\
\caption{Comparison of interpolation performance of different methods: (a) seismic data with missing traces, (b) seismic data without missing traces, (c) - (d) interpolation results of POCS, U-Net, and the proposed SFPM method, (f) - (h) residual results of POCS, U-Net, and the proposed SFPM method.}
\label{fig:fig8}
\end{figure*}

We compare the projection onto convex sets (POCS) method [\cite{cetin2013projections}], U-Net method, and the proposed SPFM to evaluate the interpolation performance of different methods. POCS, as a classic traditional interpolation method, restores missing data by an iterative projection algorithm under specific constraints, which can improve the completeness of seismic data. The interpolation results are shown in Fig. \ref{fig:fig8}. It can be observed that all methods are capable of effectively reconstructing missing seismic traces. However, the interpolation results of the POCS method exhibit noticeable interpolation boundary effects. The interpolation results of the U-Net method show poor stratigraphic continuity, and signal leakage occurs in some areas (as indicated by the red box). SPFM, while reconstructing missing seismic traces, is able to effectively recover weak signals, maintain good stratigraphic continuity, and significantly reduce signal leakage, thus outperforming other methods in terms of seismic interpolation performance and weak signal recovery.

\subsection{Seismic frequency-band extrapolation}\label{subsec7}

The low-frequency components of seismic data are of significant value in geological structure analysis, seismic tomography, and inversion, especially in oil and gas reservoir evaluation and deep imaging, where they can provide more accurate subsurface information [\cite{dong2024global}, \cite{cheng2024self}]. However, low-frequency signals are prone to attenuation or resolution limitations during seismic wave propagation, making them difficult to effectively capture, especially in deep exploration. To address this issue, we selected data from the Kerry survey area in New Zealand as the research object and extracted 1200 seismic profiles. The low-frequency components were removed from these seismic profiles by smoothing filtering to simulate low-frequency missing data, creating a low-frequency extrapolation dataset consisting of 1200 pairs of missing low-frequency/complete frequency-band, where 840 pairs were used for training, 120 for validation, and 240 for testing.

\begin{figure*}
\centering
\includegraphics[width=5in, keepaspectratio]{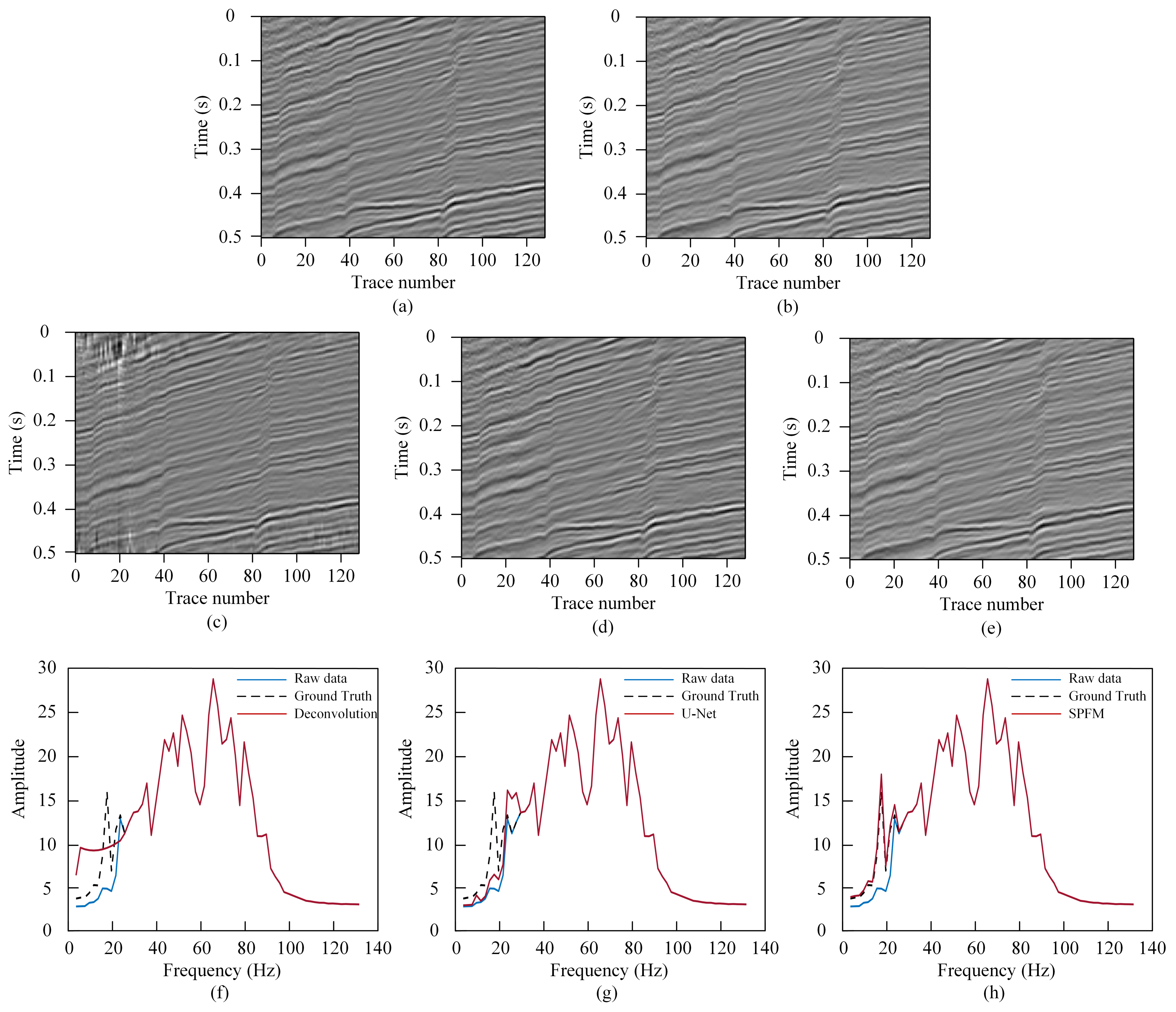}\\
\caption{Comparison of frequency-band extrapolation performance of different methods: (a) seismic data with missing low-frequency components (frequencies below 20 Hz are smoothed and filtered), (b) full-band seismic data, (c)-(e) frequency-band extrapolation results of WDM, U-Net, and the proposed SFPM method, (f)-(h) average single-trace frequency spectrum curves results of WDM, U-Net, and the proposed SFPM method.}
\label{fig:fig9}
\end{figure*}

We input the training set into the pretrained SPFM model for adaptation and obtained a model capable of performing low-frequency extrapolation. To evaluate the performance of different methods in frequency-band extrapolation, we compared SPFM with deconvolution and U-Net methods. The experimental results are shown in Fig. \ref{fig:fig9}. Both deconvolution and U-Net methods effectively compensated for the low-frequency components of the seismic data to some extent. However, the low-frequency extrapolation results obtained by deconvolution showed a decrease in SNR, with signal distortion occurring in some areas. This may be due to the accumulation of estimation errors in the low-frequency components or filtering effects during the frequency extrapolation process. The low-frequency extrapolation results obtained by U-Net method failed to effectively reconstruct the low-frequency information and instead enhanced components of other frequency bands. This could be attributed to U-Net’s insufficient global feature extraction ability, which failed to adequately capture the global information in the seismic data, leading to a loss of low-frequency components.

\subsection{Seismic resolution enhancement}\label{subsec8}

Seismic resolution is a key factor affecting the ability to finely characterize underground structures. Due to factors such as high-frequency attenuation and sparse sampling, seismic images often exhibit lower resolution, which can adversely affect subsequent seismic interpretation [\cite{zhang2022improving}, \cite{gao2023deep}, \cite{cheng2025self}]. The goal of seismic resolution enhancement is to improve the dominant frequency and broaden the signal bandwidth, particularly by improve high-frequency components, thereby better recover complex geological features in the seismic data, such as faults and folds. In this situation, we degraded the data from the Kerry survey area in New Zealand, including downsampling and low-pass filtering, to simulate the propagation characteristics of seismic waves underground and the impact of sparse sampling. By this way, we obtained a dataset of 1200 pairs of low-resolution/high-resolution paired data, with 840 pairs used for training, 120 pairs for validation, and 240 pairs for testing.

\begin{figure*}
\centering
\includegraphics[width=5in, keepaspectratio]{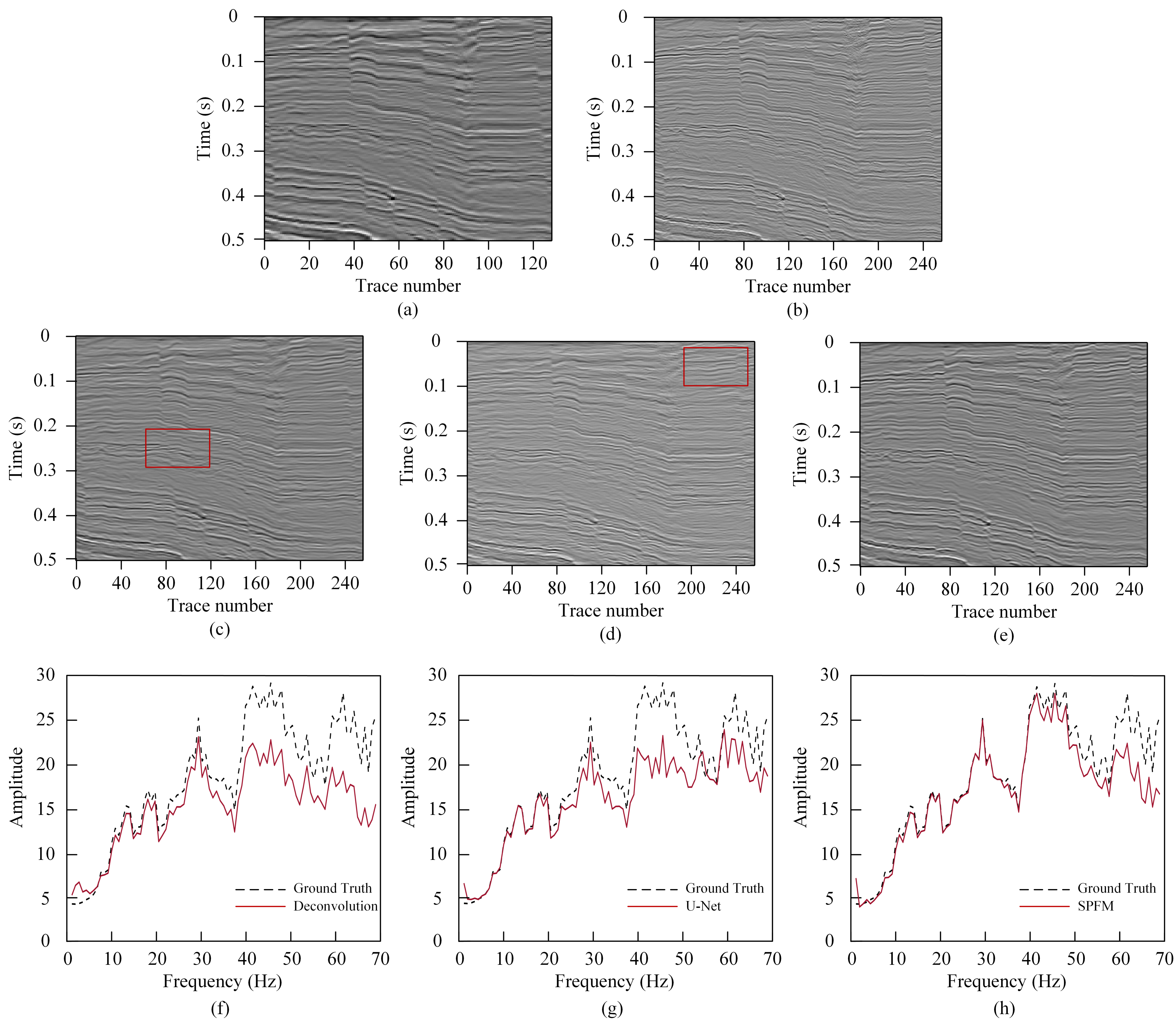}\\
\caption{Comparison of resolution enhancement performance of different methods: (a) low-resolution seismic data, (b) high-resolution seismic data, (c) - (d) resolution enhancement results of deconvolution, U-Net, and the proposed SFPM method, (f) - (h) average single-trace spectra results of deconvolution, U-Net, and the proposed SFPM method.}
\label{fig:fig10}
\end{figure*}

We fine-tuned the SPFM using the training set and compared it with the deconvolution and U-Net methods on the test set. Deconvolution is a frequency-domain signal processing technique that suppresses noise in seismic data by inverse filtering, thereby restoring high-frequency components and improving the resolution of seismic profiles. The experimental results are shown in Fig. \ref{fig:fig10}. The deconvolution method exhibits the lowest resolution and poor seismic continuity, especially in the area indicated by the red box, where significant continuity loss is evident. This issue may be due to excessive filtering of the signal during the deconvolution process. While the U-Net method enhances the resolution of seismic data in some areas, it still suffers from partial feature loss, particularly in the region marked by the red box. In contrast, the SPFM method not only significantly improves the resolution of the seismic profile but also preserves the continuity of weak signals. From the comparison of the average single-trace spectrum, it can be noticed that the enhancement results of SPFM are closer to high-resolution seismic data, demonstrating its advantages in preserving detail and enhancing resolution.

\section{Discussion}\label{sec5}

\subsection{Can the dimensionality-reduced seismic profile replace all augmented and field data?}\label{subsec9}

The proposed SPFM model is pre-trained on a non-redundant and representative dataset obtained by the augmentation, dimensionality reduction, clustering, and sampling of the field data. Therefore, the question arises whether the seismic profiles processed in this way can replace all augmented and measured data. To investigate this, we pre-train the SPFM model on both the full set of field and augmented data, and compare the performance of the pre-trained models on downstream tasks with those obtained from pre-training on the data after augmentation, dimensionality reduction, clustering, and sampling. To evaluate the model performance, we use PSNR and SSIM as evaluation metrics. The experimental results are shown in Table \ref{tab:tab2}. The model pre-trained using our proposed data augmentation strategy achieves nearly the same performance on downstream tasks as the model pre-trained on all field and augmented data, with minimal differences.

\begin{table}[h]
\caption{Performance comparison for different methods}\label{tab:tab2}%
\footnotesize
\begin{tabularx}{\textwidth}{@{} >{\centering\arraybackslash}X >{\centering\arraybackslash}X >{\centering\arraybackslash}X >{\centering\arraybackslash}X >{\centering\arraybackslash}X >{\centering\arraybackslash}X @{}}
\toprule
 & & \textbf{Denoising} & \textbf{Interpolation} & \textbf{Frequency-band\newline Extrapolation} & \textbf{Resolution\newline Enhancement} \\
\midrule
\multirow{2}{*}{Full data} & PSNR & \textbf{36.028} & 32.604 & 29.042 & \textbf{34.920} \\
 & SSIM & \textbf{0.9825} & 0.9671 & 0.8912 & \textbf{0.9851} \\
\multirow{2}{*}{Ours} & PSNR & 35.924 & \textbf{33.004} & \textbf{29.117} & 34.867 \\
 & SSIM & 0.9796 & \textbf{0.9702} & \textbf{0.8994} & 0.9741 \\
\bottomrule
\end{tabularx}
\footnotetext{Full data: the augmented data combined with the field data; Ours: the dataset filtered by our proposed strategy.}
\end{table}

\subsection{Complexity analysis}\label{subsec10}

In the field of seismic data processing, complexity analysis is crucial for the design and application of foundational models. The time complexity directly determines the training and inference speed, while the parameter complexity affects the computational resource requirements and storage overhead. The SPFM model we propose, pre-trained on a dataset constructed using our proposed dataset building strategy, requires only 19.50 hours for training. For downstream tasks, the inference time for each 128×128 input seismic profile is only 0.0275 seconds. Additionally, the SPFM has approximately 63 million parameters. Compared to foundational models in other fields and existing seismic processing models, SPFM demonstrates optimal performance in both time complexity and parameter complexity, thus validating its advantages in seismic data processing tasks.

\section{Conclusion}\label{sec}

To address the challenges posed by large data and large model parameters in foundational models within the geophysical field, we propose an intelligent seismic data processing workflow. By incorporating strategies such as data augmentation, dimensionality reduction clustering, and stratified sampling, we successfully constructed a representative and non-redundant dataset, effectively reducing the redundancy typically found in traditional larger data processing. Meanwhile, we designed a seismic processing foundational model (SPFM) based on the Mamba framework. This model not only maintains efficient seismic processing performance but also effectively avoids the quadratic complexity issues associated with the self-attention mechanism in Transformer models, significantly improving computational efficiency while preserving global feature extraction capabilities and avoiding high computational costs. By a self-supervised pretraining-fine-tuning strategy, our workflow achieved excellent results in multiple post-stack seismic data processing tasks.

However, for seismic data, 3D and even five-dimensional (5D) features are of greater importance. Existing foundational models primarily focus on processing 2D seismic profile. In the future, we plan to introduce novel transfer learning strategies to further extend the application of seismic processing foundational models to high-dimensional seismic data processing tasks.

\backmatter

\bmhead{Acknowledgements}

This work was supported in part by the National Natural Science Foundation of China under Grant 42204114, and BGP's Science and Technology project (01-04-02-2024).

\section*{Declarations}

\bmhead{Conflict of interest}

The submission of this manuscript involves no conflicts of interest, and all authors have agreed to its publication.


\bibliography{sn-bibliography}

\end{document}